\documentclass[aps,prl,twocolumn,floats,epsfig,superscriptaddress]{revtex4}
\usepackage{amsmath}
\usepackage{graphicx}
\usepackage{graphics}
\usepackage{amssymb}
\usepackage{epsfig}
\usepackage{times}

\renewcommand{\-}{\overline}

\begin{document}

\title{Quantum order by disorder in frustrated diamond lattice antiferromagnets}

\author{Jean-S\'ebastien Bernier}
\affiliation{Department of Physics, University of Toronto, Toronto, Ontario M5S 1A7, Canada}
\author{Michael J. Lawler}
\affiliation{Department of Physics, University of Toronto, Toronto, Ontario M5S 1A7, Canada}
\author{Yong Baek Kim}
\affiliation{Department of Physics, University of Toronto, Toronto, Ontario M5S 1A7, Canada}

\date{\today}

\begin{abstract}
We present a quantum theory of frustrated diamond lattice antiferromagnets. Considering quantum 
fluctuations as the predominant mechanism relieving spin frustration, we find a rich phase diagram 
comprising of six phases with coplanar spiral ordering in addition to the N\'eel phase. By computing 
the specific heat of these ordered phases, we obtain a remarkable agreement between $(k,k,0)$-spiral 
ordering and the experimental specific heat data for the diamond lattice spinel compounds
MnSc$_2$S$_4$, Co$_3$O$_4$ and CoRh$_2$O$_4$, {\it i.e. specific heat data} is a strong evidence 
for $(k,k,0)$-spiral ordering in all of these materials.
This prediction can be tested in future neutron scattering experiments on 
Co$_3$O$_4$ and CoRh$_2$O$_4$, and is consistent with existing neutron scattering data
on MnSc$_2$S$_4$. Based on this agreement we infer a monotonically 
increasing relationship between frustration and the strength of quantum fluctuations.
\end{abstract}
\maketitle

\emph{Introduction.} In insulating magnetic materials, new phases of matter may be found by letting 
local exchange interactions compete. In such situations, the spins are said to be frustrated and 
intriguing new phases such as ordered phases with coplanar or spiral ordering or ``spin liquid''
paramagnets can arise.\cite{moessner} In any given frustrated material, the ground state may be 
determined by identifying the primary mechanism relieving the frustration. While extrinsic mechanisms, 
such as small dipole interactions, disorder or lattice distortions\cite{pyc}, 
may be important, perhaps the most interesting possibility is when temperature or quantum fluctuations 
alone relieve the frustration, a process termed ``order by disorder''.\cite{vilshend}  

In this light, recent experiments which unveil strong frustration in spinel compounds
AB$_2$X$_4$, with magnetic ions occupying the A-sites, are particularly interesting.
Here the A-sites form a diamond lattice of spin $S=\frac{3}{2}, 2, \frac{5}{2}$ local moments.
Important examples include seven diamond spinels (\cite{exp} and references within), four 
that order (MnSc$_2$S$_4$, MnAl$_2$O$_4$, Co$_3$O$_4$, CoRh$_2$O$_4$) and three that do not 
down to the lowest temperatures studied (CoAl$_2$O$_4$, FeAl$_2$O$_4$, FeSc$_2$S$_4$).
When the moments order, the ordering temperature,
$T_c$, is low compared to the Curie-Weiss temperature, $\Theta_{\text{CW}}$, with 
frustration parameters\cite{ramirez}, $f=\frac{|\Theta_{\text{CW}}|}{T_c}$, varying from $1.2$ to $10$. 
The magnetic ordering in one of the ordered materials, MnSc$_2$S$_4$, has been identified
as an exotic $(k,k,0)$ coplanar spiral via extensive neutron scattering experiments\cite{exp}
while the magnetic ordering patterns of other ordered diamond spinels are not determined yet. 
Given that a diamond lattice is bipartite, this ubiquitous evidence for frustration is highly unexpected.

In combination with their frustrated magnetic properties, 
diamond spinels also have unusual temperature dependence of specific heat.
Remarkably, among the four materials that order, their specific heat data share the same unusual behavior 
below $T_c$ (see Ref.\cite{exp}). Instead of a pure $T^3$ power-law expected for 
incommensurate magnetic ordering in three dimensions, two inflection points are observed.  
The three that do not order also share the same 
characteristic specific heat, but is quite different from those that order. These materials 
display a $T^{2.5}$ power-law over a decade in temperature.\cite{exp} 

Following these experimental discoveries, the classical Heisenberg model on the diamond
lattice with the nearest (nn) and next-nearest neighbor (nnn) exchange interactions has been
studied\cite{bergman}. It was demonstrated that the frustration arises from the nnn 
interactions that couple spins within each of the two face-centered cubic
(fcc) sublattices of the diamond lattice structure\cite{bergman}. 
This coupling creates a highly degenerate set of classical coplanar spirals whose propagation vectors 
form a continuous surface in momentum space. Relieving this classical ground state degeneracy 
by thermal fluctuations was then found to produce a rich phase diagram at the classical 
level\cite{bergman}, including the $(k,k,0)$ 
spiral phase discovered in the neutron scattering experiments on MnSc$_2$S$_4$.
This put diamond spinels in a promising class of materials 
in which (thermal) ``order by disorder'' may be experimentally observed. However, this classical
picture may not be sufficient to describe possible effect of quantum fluctuations in these materials
with relatively small spin $S=\frac{3}{2}, 2, \frac{5}{2}$ and the specific heat data below $T_c$.

In this letter, we present a quantum theory of frustrated diamond lattice antiferromagnets. 
We find that quantum fluctuations act as an order by disorder mechanism to produce a similar 
but richer phase diagram compared with that obtained exclusively from thermal fluctuations.  
In particular, focusing on the ordered states, we demonstrate that the characteristic signatures 
in the specific heat data of three of the ordered materials (MnSc$_2$S$_4$, Co$_3$O$_4$ and CoRh$_2$O$_4$) 
can be explained if the magnetic ordering pattern in these materials is $(k,k,0)$-spiral
selected by strong quantum fluctuations. 
Thus, we argue that specific heat data is a convincing evidence for $(k,k,0)$-spiral ordering
in all these materials. Based on this comparison, we show how frustration and quantum fluctuations 
are strongly intertwined in these systems. Finally, we discuss the implications of our results 
on future neutron scattering experiments on these materials.

\begin{figure}[t]
\vspace{0cm}
\includegraphics[scale=1.0]{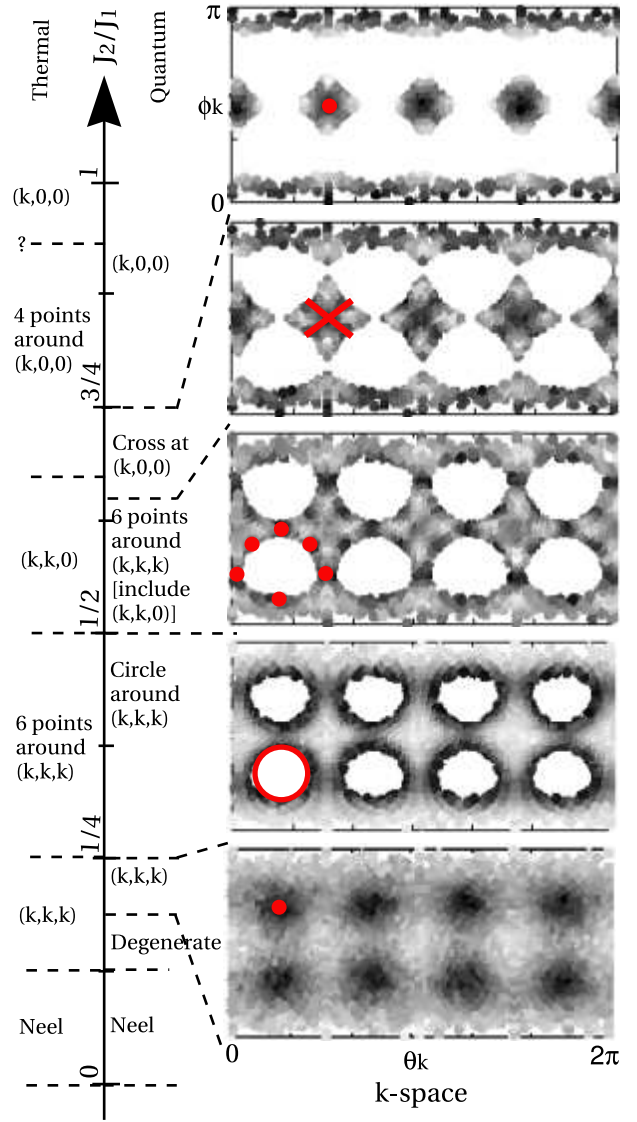}
\caption{Structure of the ground states as a function of $\frac{J_2}{J_1}$ after the inclusion 
of quantum fluctuations. The $\phi_k$ versus $\theta_k$ plots on the right are planar views 
of the somewhat distorted and/or punctured sphere forming the set of degenerate classical 
ground states. $\phi_k$ and $\theta_k$ are the polar and azimuthal angles spanning all 
directions in momentum space. Darker points represent lower energy. On each plot, 
red dots or lines give examples of states selected by quantum fluctuations. To contrast 
the effects of quantum and 
thermal fluctuations, states selected by entropic effects are summarized on the 
left.\cite{bergman}}
\label{qfgsl}
\end{figure}

Keeping in mind that frustration arises from the nnn exchange interactions\cite{bergman},
we begin with the quantum Heisenberg model:
\begin{eqnarray}
H &=& J_1\sum_{\langle ij \rangle} {\bf S}_i\cdot{\bf S}_j
+J_2 \sum_{\langle\langle ij \rangle \rangle} {\bf S}_i\cdot{\bf S}_j,
\label{hafm}
\end{eqnarray}
where ${\bf S}_i$ are spin-$S$ operators at site i, $J_1>0$ is the exchange coupling on the 
nn links (between sites on different fcc sublattices) and $J_2>0$ is the
exchange coupling on the nnn links (between sites on the same fcc sublattice). 
By studying the large-$N$ limit of the Sp($N$) 
generalization of this model\cite{sachdevread}, 
we study the role of quantum fluctuations as a controlled expansion in $\frac{1}{N}$. 
The advantage of this method is that, unlike the large-$S$ expansion\cite{henley}, the results are
non-perturbative in the spin magnitude $S$ (strength of quantum fluctuations) and  
the coupling constants ($J_1$ and $J_2$).
The resulting phase diagram, which indeed exhibits order 
by disorder, is presented in Fig.\ref{qfgsl}. For $\frac{J_2}{J_1}>1/8$, the energy of the degenerate 
set of classical states is given by the grey-to-black pixellated surface on the right hand side 
of the figure. Zero point energy corrections due to quantum fluctuations lift this degeneracy 
(grey-to-black pixels represent higher-to-lower energy spin configurations)
and only the black pixels with the smallest energy remain degenerate. The phase diagram was
obtained based on the resulting magnetic order of the selected ground state.

Fig.\ref{qfgsl} also shows that quantum and thermal fluctuations lift the classical 
ground state degeneracy differently. While thermal fluctuations entropically select an ordering pattern, 
quantum fluctuations select an ordering configuration purely from energetic considerations and so need 
not select the same state.
Comparing our results with a study where only entropic effects were 
considered\cite{bergman}, we notice that both kinds of fluctuations select states along similar 
high symmetry directions such as $(k,k,k)$, $(k,k,0)$ and $(k,0,0)$. However, similar states are 
not always present in the same range of $\frac{J_2}{J_1}$. In addition, within numerical
accuracy, quantum fluctuations do not always lift the degeneracy, or lift it only partly 
as in the ``degenerate'', ``circle'' and ``cross'' phases. This is in contrast to thermal fluctuations, 
where points of lowest energy are always selected when entropic effects are considered. 

\begin{figure}[t]
\vspace{0cm}
\rotatebox{270}{
\includegraphics[scale=0.35]{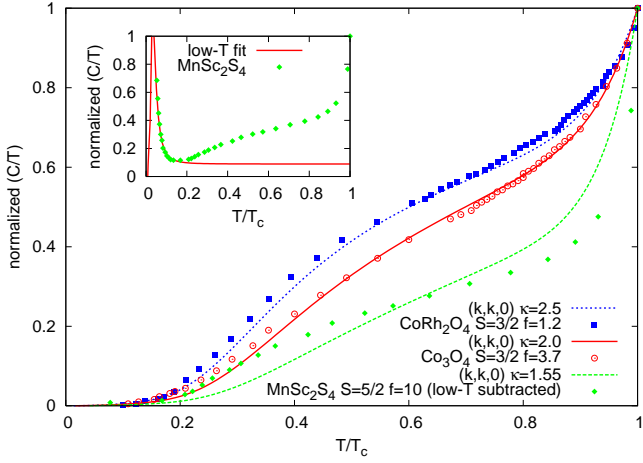}}
\caption{Comparison of specific heat data of CoRh$_2$O$_4$, Co$_3$O$_4$, MnSc$_2$S$_4$ and 
the theoretical large-$N$ specific heat of the $(k,k,0)$ spiral ordering with $J_2/J_1=0.6$. 
Here $1/\kappa=N/2S_\text{eff}$, 
held fixed in the large-$N$ limit, gives the strength of quantum fluctuations and increases monotonically 
with the frustration parameter $f = |\Theta_\text{CW}|/T_c$. The inset shows the subtraction of the 
nuclear contribution with a constant ``background'' from the specific heat of MnSc$_2$S$_4$ 
using $C_I/T = \Delta^2/T^3 (e^{(\Delta/T)}/(1+e^{(\Delta/T)})^2 + C_0$.}
\label{sh}
\end{figure}

Focusing on the diamond spinels that magnetically order at low temperatures, we find remarkable 
agreement between the measured heat capacity and our large-$N$ theory in a phase with $(k,k,0)$ ordering. 
As shown in Fig.\ref{sh}, $\frac{C}{T}$ for CoRh$_2$O$_4$, Co$_3$O$_4$ and 
MnSc$_2$S$_4$, have two characteristic inflection points before reaching $T_c$, a feature that is best 
reproduced by the $(k,k,0)$ spiral ordering in the large-$N$ limit. 
Consequently, we propose that all of these spinels most likely have $(k,k,0)$ spiral ordering. 
This result is in agreement with neutron scattering experiments on 
MnSc$_2$S$_4$.\cite{exp} Future neutron scattering experiments on CoRh$_2$O$_4$ and Co$_3$O$_4$ could 
verify this 
prediction.

Through the remarkable fitting of the large-$N$ theory to the heat capacity data, we also find a simple 
relationship between the empirical frustration parameter, $f$, and 
the strength of quantum fluctuations given by the large-$N$ effective spin length parameter 
$\kappa = \frac{2S_\text{eff}}{N}$ (here $\kappa$ is held fixed in the $N \rightarrow \infty$ limit). 
To describe adequatly the experimental specific heat data of the more 
frustrated (larger $f$) compounds, we need to include stronger 
quantum fluctuations than required to describe the 
moderately frustrated ones. As shown in Fig.\ref{sh}, the spin $\frac{5}{2}$ system MnSc$_2$S$_4$ 
with frustration parameter $f=10$ is best fitted using $\kappa=1.55$ while the spin $\frac{3}{2}$ system 
CoRh$_2$O$_4$ with $f=1.2$ is best fitted with $\kappa=2.5$.  Thus for the diamond spinels that 
order at low temperatures, quantum fluctuations correlate with frustration much more than 
with the physical spin representation.

\emph{General Sp($N$) mean-field free energy.}  The Hamiltonian describing the interactions between 
spins on the diamond lattice is given by Eq.(\ref{hafm}). The SU($2$) spin symmetry of the Hamiltonian is
generalized to Sp($N$) by first recasting the spins using the bosonic representation 
${\vec S}_i = \frac{1}{2} b^{\dagger}_{i \alpha} {\vec \sigma}_{\alpha \beta} b_{i\beta}$ 
where $\alpha, \beta =\{\uparrow,\downarrow\}$ labels two possible spin states of each boson 
and then by introducing $N$ flavors of such bosons on each site. In order to keep the physical Hilbert
space of spins, a constraint on the number of bosons given by 
$n_{b} = b^{\dagger m}_{i \alpha} b^{m}_{i\alpha} = 2S_\text{eff} = \kappa N$ where $m = 1,...,N$ must 
be imposed at each site. Note that $N=1$ corresponds to the physical limit Sp(1) $\equiv$ SU(2).
The action of the Sp($N$) generalized model is then given by
\begin{eqnarray}
\mathcal{S} = \int_0^\beta d\tau \{\-{b}_{i \alpha}^{m} \partial_{\tau} b_{i\alpha}^{m}
-\frac{J_{ij}}{2 N} \-{A}_{ij} A_{ij} + \lambda_i (\-{b}_{i\alpha}^m b_{i\alpha}^m - n_{bi}) \}
\nonumber
\end{eqnarray}
where $A_{ij} = \epsilon_{\alpha \beta}~\delta_{m m'} b_{i\alpha}^{m} b_{j\beta}^{m'}$
($\epsilon_{\alpha\beta}~\delta_{m m'}$ is the Sp($N$) generalized antisymmetric tensor of SU($2$)) 
and the chemical potential $\lambda_i$ keeps the average number of bosons fixed to
$n_b = \kappa N$ at every site. The mean-field action is then obtained by decoupling the quartic boson
interaction in $\mathcal{S}$ using the Hubbard-Stratonovitch fields $Q_{ij} = -Q_{ji}$ directed
along the lattice links so that one obtains $Q_{ij} = \langle A_{ij} \rangle / N$ at the saddle point. 
The mean field solution becomes exact in the large-$N$ limit where $N \rightarrow \infty$ is
taken while $\kappa = n_b/N$ is fixed.
We also introduce the paramerization 
$b_{i\alpha}^{m} = \left( \begin{array}{c} \sqrt{N} x_{i\alpha}~~b_{i\alpha}^{\tilde{m}} \end{array} \right)^T$ 
where $\tilde{m}=2,...,N$ to allow for the possiblity of long-range order that occurs when $x_{i\alpha} \not= 0$. 
Consequently, after integrating over the bosons, and rescaling 
$Q_{ij}$ and $\lambda$ by $\kappa$, $x_{i\alpha}$ by $\sqrt{\kappa}$ and the temperature by $\kappa^2$,
we obtain the mean-field free energy
\begin{eqnarray}
\frac{F}{N\kappa^2} &=& \sum_{i,j} \frac{J_{ij}}{2} (|Q_{ij}|^2 
- Q_{ij} (\epsilon_{\alpha\beta} x^*_{i\alpha} x^*_{j\beta}) + c.c.) \nonumber \\ 
&+& \lambda \sum_i (|x_{i\alpha}|^2 - (\frac{1}{\kappa}+1))
+ f_{\text{eff}}
\label{ferescaled}
\end{eqnarray}
where $f_\text{eff} = \sum_\mu \frac{\omega_\mu(Q,\lambda)}{\kappa}
+2 k_B T \ln(1-e^{-\frac{\omega_\mu(Q,\lambda)}{\kappa k_B T}})$, and
$\omega_{\mu} (Q,\lambda)$ are the eigenvalues of the mean-field Hamiltonian.
Note that the chemical potential is now taken to be uniform since each site has the same number 
of nn and nnn links. In general, magnetic ordering $x_{i \alpha} \not= 0$ occurs in the 
semiclassical limit at larger $\kappa$ while quantum paramagnetic phases are obtained when $\kappa$ is small.

\emph{Classical ground state.} In the classical limit $\kappa \rightarrow \infty$ at $T\to0$, 
one can show that $Q^c_{ij} = \epsilon_{\alpha\beta}x^c_{i\alpha}x^c_{j\beta}$, so that 
the classical energy is given by
\begin{eqnarray}
\frac{E^c}{N \kappa^2}= - \sum_{i,j} 
\frac{J_{ij}}{2}|\epsilon_{\alpha\beta}x^c_{i\alpha}x^c_{j\beta}|^2 +  \lambda^c \sum_i (|x^c_{i\alpha}|^2 - 1).
\label{eclass}
\end{eqnarray}
Minimizing Eq.(\ref{eclass}) with respect to $x^c_{i\alpha}$ and $\lambda^c$ is equivalent 
to determining the classical ground states of Eq.(\ref{hafm}) provided the solution has $|x^c_{i\alpha}|^2=1$. 
Rewriting $E^c$ in terms of a quadratic form in the classical unit spin vectors
$\vec S_i^c = x^{c*}_{i\alpha}\vec\sigma_{\alpha\beta}x^c_{i\beta}$
and transforming to {\bf k}-space unveils two bands
$\epsilon_{\pm}({\bf k}) = J_2(\Lambda^2({\bf k})-1)\pm \tfrac{1}{2}J_1\Lambda({\bf k})$
where $\Lambda^2({\bf k}) = 4\{\prod_{u=x,y,z}\cos^2\frac{k_u}{4}
+\prod_{u=x,y,z}\sin^2\frac{k_u}{4}\}$. The minimum eigenvalue is obtained in the lower 
band $\epsilon_{-}$ and is unique at ${\bf k}=0$ for $\frac{J_2}{J_1}<\frac{1}{8}$ but highly degenerate,
corresponding to a surface in {\bf k}-space, for $\frac{J_2}{J_1}\geq \frac{1}{8}$. 
For $\frac{1}{8} \leq \frac{J_2}{J_1}\leq \frac{1}{4}$,  
the space of ground states resembles the surface of a slightly deformed sphere. For larger $\frac{J_2}{J_1}$, 
the size of the ``sphere'' increases and eight holes centered around the $(k,k,k)$ directions begin 
to puncture its surface (see also Ref.\cite{bergman}). The solution of the classical limit is then 
completed by finding for each degenerate eigenstate, 
labeled by its ${\bf k}$ value, its real space spin configuration and chemical 
potential. In terms of our spinor representation, we obtain
$x^{c,A}_{i\uparrow/\downarrow} = \pm \tfrac{1}{\sqrt{2}} e^{\mp \tfrac{i}{2}({\bf k}\cdot{\bf r}_i-\tfrac{1}{2}\theta({\bf k}))}$ and
$x^{c,B}_{i\uparrow/\downarrow} = \frac{1}{\sqrt{2}} e^{\mp \tfrac{i}{2}({\bf k}\cdot{\bf r}_i+\tfrac{1}{2}\theta({\bf k}))}$
where A and B label the two fcc sublattices and 
$\theta({\bf k}) = \arctan(\tan\frac{k_x}{4}\tan\frac{k_y}{4}\tan\frac{k_z}{4})$. 
Varying $E^c$ with respect to $x^c_{i\alpha}$ gives the chemical potential
$\lambda^c_{i}=\frac{1}{x^{c *}_{i\uparrow}}\sum_j\frac{J_{ij}}{2}Q^{c *}_{ij}x^c_{j\downarrow}$.

\emph{Effect of quantum fluctuations.} Given the above solution to the classical ground states, 
consider expanding the ground state energy $E$ in powers of $1/\kappa$ so that 
$E = E^c + \frac{1}{\kappa} E^1 + \ldots$. This leads to the quantum correction 
\begin{eqnarray}
\frac{E^1}{N\kappa^2} &=& \sum_\mu \omega_\mu(Q^c,\lambda^c)- \lambda^c N_S
\nonumber
\end{eqnarray}
where $N_S$ is the number of lattice sites and $Q^c$, the classical values for the link
variables, are given by $Q^{c AA/BB}_{ij} = \mp i\sin(\tfrac{{\bf k}\cdot({\bf r}_j-{\bf r}_i)}{2})$
and $Q^{c AB/BA}_{ij} = \pm \cos(\tfrac{{\bf k}\cdot({\bf r}_j-{\bf r}_i) \pm \theta({\bf k})}{2})$.
Since $Q^c_{ij}$ is only dependent on the difference between two sites,
we can Fourier transform 
back to {\bf k}-space and solve for $E^1$ analytically as a sum over wave-vectors. In practice, 
this energy correction is computed using an adaptive Monte Carlo integration method and 
the new ground state is found by sampling the surface of equal energy. The first order quantum 
corrections dramatically alter the topology of the degenerate ground state manifold reducing 
the ${\bf k}$-space surface of lowest energy to only points or lines
represented by the black pixels in Fig.\ref{qfgsl}. For $0.125<\frac{J_2}{J_1}<0.18$, 
the ``sphere'' of equal energy remains surprisingly degenerate; 
for $0.18<\frac{J_2}{J_1}<0.25$, the eight $(k,k,k)$ directions are selected; 
for $0.25<\frac{J_2}{J_1}<0.5$, states labeled by ${\bf k}$ points forming eight circles 
around the $(k,k,k)$ directions are chosen; for $0.5<\frac{J_2}{J_1}<0.65$, each circle 
gives way to six points around each $(k,k,k)$ direction (among which three points 
are $(k,k,0)$ directions); for $0.65<\frac{J_2}{J_1}<0.75$, the states labeled by ${\bf k}$ 
points form a degenerate cross centered around each $(k,0,0)$ direction and finally 
for $\frac{J_2}{J_1}>0.75$ states labeled by points along the six $(k,0,0)$ directions become 
the states of lowest energy.  

\emph{Specific heat and comparison to experiments.} To compare the above order by disorder 
predictions with current experiments, we compute the specific heat. 
At finite temperatures, we assume that the phase and amplitude of both bonds 
and condensates can vary but that these changes are spatially uniform. 
We write $Q_{ij} = R~e^{i\xi}~Q^c_{ij}$ and 
$x_{i\alpha} = \sqrt{r}~e^{i\zeta}~x^c_{i\alpha}$
where $Q^c_{ij}$ and $x^c_{i\alpha}$ are bond and condensate values in one of the
$T=0$ spin configurations chosen by quantum fluctuations. For a given $T$
and effective spin length, $\kappa$, $R(Q,\lambda,T)$, $r(Q,\lambda,T)$, $\xi$ and $\zeta$ 
are obtained from self-consistent saddle point equations.
For spiral configurations $(k,k,k)$, $(k,k,0)$ and $(k,0,0)$,
the ground state is magnetically ordered at $T=0$ if $\kappa > \kappa_c$
where, for example, $\kappa_c = 0.17, 0.094, 0.087$ for $\frac{J_2}{J_1} = 0.2, 0.6, 0.85$.

Using $Q_{ij}$ and $x_{i\alpha}$ in Eq.(\ref{ferescaled}), 
we obtain the specific heat $C=-T \frac{\partial^{2}F}{\partial T^2}$. In 
the limit of very low temperatures, $R$ and $r$ are approximately $T$-independent
and $C \sim T^3$ as expected for 3D antiferromagnets. However,
as temperature is increased, $R$ and $r$ become $T$-dependent and 
the specific heat departs from its $T^3$ behavior. 
As shown in Fig.\ref{sh}, the specific heat obtained from the 
$(k,k,0)$ spiral ordering with $\frac{J_2}{J_1}=0.6$ agrees well with the experimental specific heat data
of CoRh$_2$O$_4$ and Co$_3$O$_4$ and the fit looks reasonable for MnSc$_2$S$_4$.
For these three materials, $\frac{C}{T}$ presents two inflection points before $T_c$, a 
feature that is best reproduced by $(k,k,0)$ ordering and is robust
for a finite range of $\frac{J_2}{J_1}$ as long as the magnetic ordering remains 
the same. Other ordering wavevectors do not reproduce as nicely the 
characteristic temperature dependence of the specific heat data.
The two inflection points arise due to the temperature dependence of the magnon spectrum. The
first point arises as the spin-wave velocity departs from its $T=0$ value
while the second point appears when the temperature becomes of the order of the
magnon bandwidth.
Regarding MnSc$_2$S$_4$, the low temperature part of the experimental specific heat
has substantial nuclear spin contributions that we subtract 
using a simple two-level system formula as described in the figure caption 
of Fig.\ref{sh}. Also, Ref.\cite{bergman} pointed out that the neutron scattering
experiment on this material suggests $\frac{J_2}{J_1}=0.85$ and implies that 
a third neighbor coupling $J_3$ is necessary to stabilize $(k,k,0)$ order for
this value of $\frac{J_2}{J_1}$. We expect that the latter and the error arising from the subtraction 
of the nuclear contribution are the reasons behind the less satisfactory fit for MnSc$_2$S$_4$.

\emph{Conclusion.} 
We presented a theory of frustrated diamond lattice quantum 
antiferromagnets. Considering quantum fluctuations as the predominant mechanism relieving spin 
frustration in this spin system we found a rich phase diagram consisting of six phases 
with coplanar spiral ordering in addition to the N\'eel phase. By comparing specific heat curves 
found in the large-$N$ mean-field theory with data obtained from 
CoRh$_2$O$_4$, Co$_3$O$_4$ and MnSc$_2$S$_4$, 
we propose that they all share the same magnetic order 
in their ground state: a coplanar spiral with propagation vector $(k,k,0)$. 
Note that the neutron scattering data is currently available only for
MnSc$_2$S$_4$.
From the fit in Fig.\ref{sh}, we conclude that a remarkable correlation exists 
between the strength of quantum fluctuations, measured by the 
effective spin magnitude $\kappa = \frac{2S_\text{eff}}{N}$, and the empirical frustration 
parameter $f = \frac{|\Theta_{\text{CW}}|}{T_c}$ while $\kappa$ and the physical spin 
magnitude $S=\frac{3}{2}, \frac{5}{2}$ of the magnetic ions appear only loosely related. 
We expect future neutron scattering experiments will verify our predictions. 

We thank J. Alicea, D. Bergman, L. Balents, S. Takei and S. Trebst for 
helpful discussions. This work was supported by 
NSERC, CRC, CIFAR, KRF-2005-070-C00044 (JSB, MJL, YBK), and FQRNT and OGS (JSB).

\end{document}